\documentclass[]{spie}  

\usepackage{amsmath,amsfonts,amssymb}
\usepackage{graphicx}
\usepackage[colorlinks=true, allcolors=blue]{hyperref}

\title{Development of an ultra-sensitive 210-micron array of KIDs for far-IR astronomy}

\author[a]{Elijah Kane}
\author[a]{Chris Albert}
\author[c]{Nicholas Cothard}
\author[a]{Steven Hailey-Dunsheath}
\author[b]{Pierre Echternach}
\author[a]{Logan Foote}
\author[b]{Reinier M. Janssen}
\author[b]{Henry (Rick) LeDuc}
\author[a]{Lun-Jun (Simon) Liu}
\author[b]{Hien Nguyen}
\author[c]{Jason Glenn}
\author[b]{Charles (Matt) Bradford}
\author[a]{Jonas Zmuidzinas}
\affil[a]{California Institute of Technology, 1200 E California Blvd, Pasadena, 91125, California, USA}
\affil[b]{Jet Propulsion Laboratory, California Institute of Technology, 4800 Oak Grove Dr, Pasadena, 91109, California, USA}
\affil[c]{NASA Goddard Space Flight Center, 8800 Greenbelt Rd, Greenbelt, 20771, Maryland, USA}

\authorinfo{Further author information: (Send correspondence to Elijah Kane)\\Elijah Kane: E-mail: ekane@caltech.edu}

\begin{document} 
\maketitle

\begin{abstract}

The Probe far-Infrared Mission for Astrophysics (PRIMA) is a proposed space observatory which will use arrays of thousands of kinetic inductance detectors (KIDs) to perform low- and moderate-resolution spectroscopy throughout the far-infrared. The detectors must have noise equivalent powers (NEPs) at or below $1\times 10^{-19}$~W~Hz$^{-1/2}$ to be subdominant to noise from sky backgrounds and thermal noise from PRIMA's cryogenically cooled primary mirror. Using a Radio Frequency System on a Chip for multitone readout, we measure the NEPs of detectors on a flight-like array designed to observe at a wavelength of 210~$\mu$m. We find that $92\%$ of the KIDs measured have an NEP below $1\times 10^{-19}$ W Hz$^{-1/2}$ at a noise frequency of 10 Hz.
\end{abstract}


\section{INTRODUCTION}
\label{sec:intro}  

The far-infrared is rich in astrophysical information about dusty galaxies, as interstellar dust absorbs visible and ultraviolet light and re-radiates in the far-infrared. In addition to the continuum emission of dust, ionized atoms have fine-structure emission lines which can trace energetic processes in galaxies such as star formation, active galactic nucleus accretion, and feedback. 

The Probe far-Infrared Mission for Astrophysics (PRIMA) is a space probe concept aiming to observe in the far-IR from Earth-Sun L2 orbit. PRIMA's Far-Infrared Enhanced Survey Spectrometer (FIRESS) will perform spectroscopic observations across the $25-235\,\mu$m range. FIRESS will have a low-resolution mode covering this full band with a spectral resolution of $R\sim 100$. FIRESS will also have a high-resolution mode with a tunable spectral resolution, with a maximum of $R=17,000$ at 25~$\mu$m and $R=4,400$ at 112~$\mu$m. Due to PRIMA's cryogenically cooled primary mirror, the dominant noise background seen by the detectors will be photon shot noise from zodiacal dust. To be background noise-limited, the detectors must achieve an NEP at or below  $1\times 10^{-19}$ W Hz$^{-1/2}$.

PRIMA will use kinetic inductance detectors (KIDs) to achieve this sensitivity goal. Past work for PRIMA has demonstrated individual detector NEPs below $1\times 10^{-19}$ W Hz$^{-1/2}$ \cite{shd2023}. Further work has applied a Radio Frequency System-on-a-Chip (RFSoC) multitone readout system to measure NEPs for a fraction of KIDs on a flight-like array \cite{footeltd20}. In this paper, we increase the readout yield using the RFSoC readout system and obtain NEP measurements of the majority of pixels on an array designed to observe at a wavelength of 210 $\mu$m for FIRESS.

\section{Methods}

\subsection{Device}

\begin{figure} [ht]
   \begin{center}
   \begin{tabular}{c} 
   \includegraphics[height=5cm]{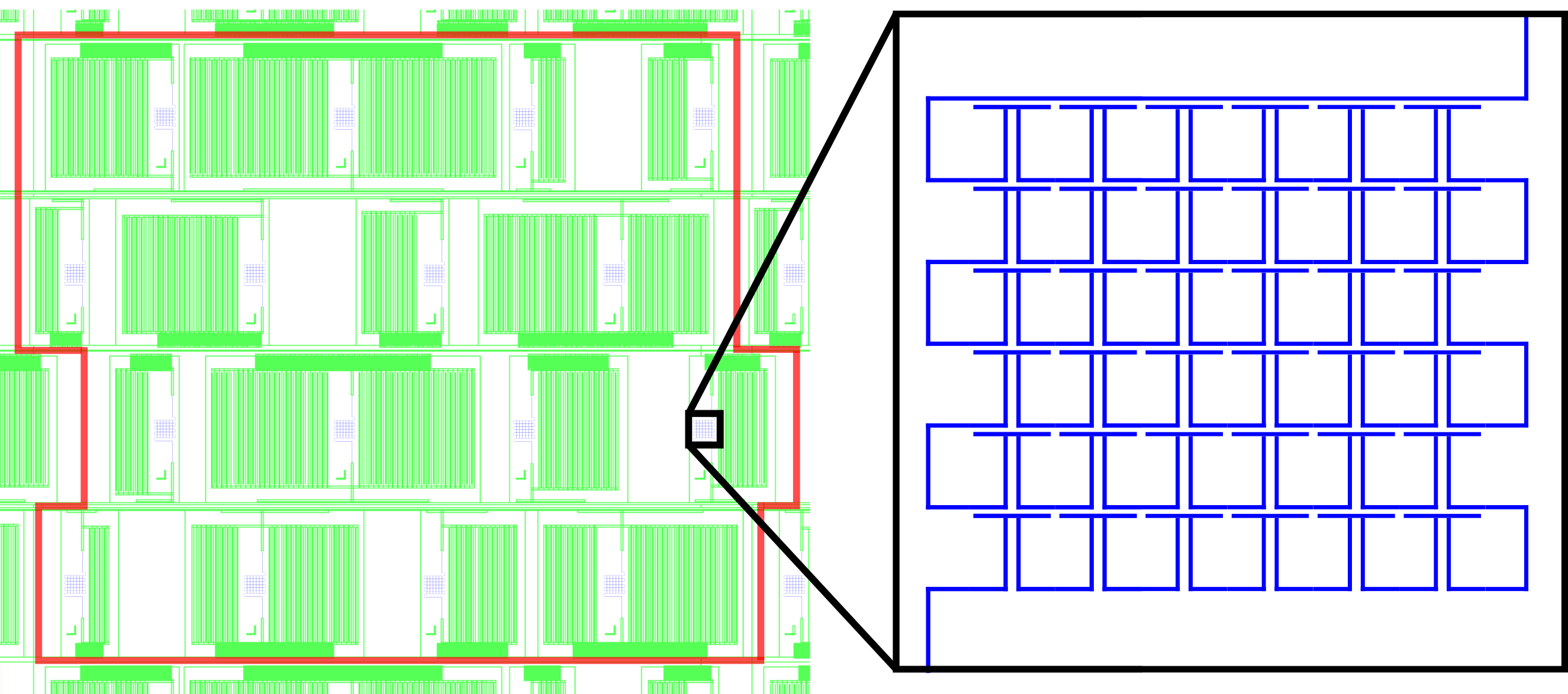}
   \end{tabular}
   \end{center}
   \caption[example] 
   { \label{fig:unitcell} 
\textit{Left:} The 4 by 4 pixel unit cell. Each detector has an aluminum absorber (blue), one or two niobium (green) main IDC capacitors, an IDC capacitor to ground, and a parallel-plate capacitor to the readout coplanar waveguide (CPW) feedline. \textit{Right:} The $\Pi$-shaped inductor/absorber design. The gaps between lines have been exaggerated for clarity.}
   \end{figure}

The device under test is a 12 by 84 pixel array. Each pixel uses the $\Pi$-shaped inductor/absorber design presented by Hailey-Dunsheath et. al. \cite{shd2023}, but with a smaller $6\times 6$ footprint yielding a lower volume of $15.6\,\mu$m. The capacitors use an interdigitated (IDC) design, and the capacitances are chosen by varying the number of tines according to a 16 KID (4 by 4) unit cell approach as described by Foote et. al. \cite{footeltd20}. The unit cell and inductor/absorber design are illustrated in Fig.~\ref{fig:unitcell}. The capacitance values have been decreased to account for the higher-than-expected inductance of the absorber observed by Foote et. al. In the finished array, the backside will be hybridized to an array of silicon microlenses \cite{cothard_monolithic_2024}. For the measurements presented in this paper, the detectors did not have microlenses.

\subsection{Measurement and analysis methods}

The device is cooled to a temperature of 125~mK using a $^3$He/$^4$He dilution refrigerator. For optical measurements, the device is illuminated with a blackbody source which can be swept from a temperature of $\sim4$~K up to 30~K. Optical filters are used to define a bandpass between the blackbody and the detectors, centered at 1.42~THz with a FWHM of 0.17~THz. The power incident on each detector is calculated to be 10~zW to 400~aW for these temperatures based on modeling of the filters and apertures between the blackbody and the detectors.

Single-tone measurements were performed using the system described in Hailey-Dunsheath et. al. \cite{shd2023} With the blackbody at its lowest setting (effectively ``dark"), we use the single-tone system to do a full transmission vs. frequency sweep of the device to identify resonances and generate a tone list for multitone measurements.

A Xilinx Radio Frequency System on a Chip (RFSoC) with firmware and software developed for the CCAT-P Prime-Cam instrument \cite{burgoyne_CCAT_2024} was used to take multitone measurements of this device. Since the digitization rate of the RFSoC's output waveform is 4 GHz, its Nyquist frequency is 2 GHz. However, there were resonant frequencies on the array up to 2.9 GHz. To avoid aliasing of resonant features from the second Nyquist zone back into the first Nyquist zone, we put a lowpass filter with a 3 dB point near 2 GHz on the output of the RFSoC's DAC port and only read out resonances below 1.9 GHz. We note that RFSoCs with higher digitization rates of 6.55~GHz and 9.85~GHz are commercially available, which would eliminate the aliasing issue, and PRIMA will use a higher digitization rate for deployment.

The readout tone power optimization is performed through an iterative process of fitting $S_{21}$ versus frequency data to a resonance model. The resonance model accounts for amplifier gain, cable delay, kinetic inductance nonlinearity \cite{Swenson_2013}, and impedance mismatch on either side of the readout feedline \cite{Khalil_2012}:

\begin{equation}
    \label{eq:s21}
    S_{21} = A(f)e^{2\pi j f \tau }\left(1-\frac{1}{1+2jy}\frac{Q_r}{Q_c\cos\phi}e^{j\phi}\right).
\end{equation}

\begin{figure} [ht]
   \begin{center}
   \begin{tabular}{c} 
   \includegraphics[height=5cm]{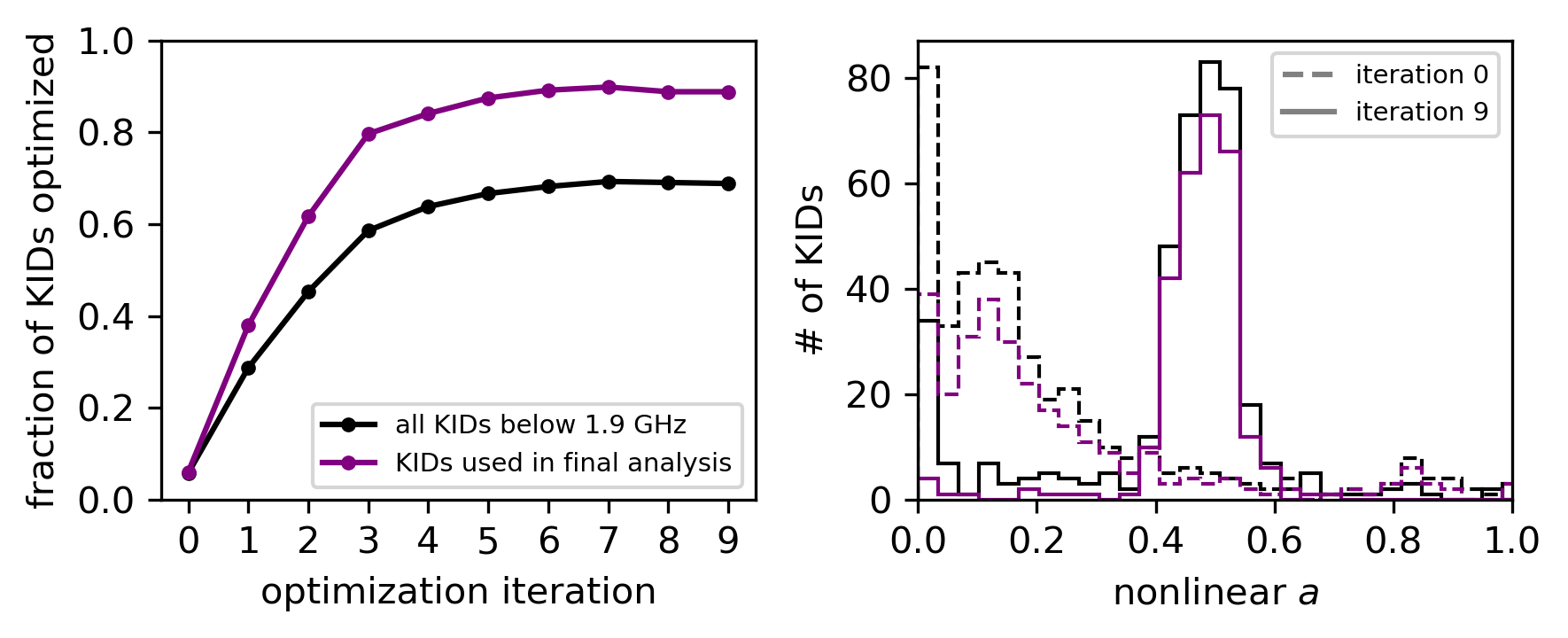}
   \end{tabular}
   \end{center}
   \caption[example] 
   { \label{fig:power_opt}
\textit{Left:} Fraction of KIDs optimized at each iteration of the power optimization procedure, where optimization is defined as $a\in[0.4,0.6].$ \textit{Right:} Histogram of fitted $a$ values at the start and end of optimization. The black lines include all KIDs below 1.9 GHz, and the purple lines include only KIDs that were used in the final noise analysis of Section~\ref{sec:results}.}
   \end{figure}

Here, $y = Q_r x + a/(1+4y^2)$ and $x = (f-f_r)/f_r$, where $a$ is the KID's nonlinearity parameter which is proportional to the readout tone power $P_g$, and $f_r$ is the resonant frequency. Our optimization goal is to read out each detector at a tone power where $a\in [0.4,0.6]$; this ensures that the frequency response of the detector is well above the amplifier noise while keeping below bifurcation $(a=0.77)$. We iteratively take sweep data, fit to Equation~(\ref{eq:s21}), and adjust each tone power by a constant multiplicative factor until the number of optimized detectors stabilizes. The convergence is also sped up by multiplying $P_g$ by $0.5/a$ if the fitted $a$ is above $0.1$. The results of the optimization procedure are shown in Fig.~\ref{fig:power_opt}, demonstrating rapid convergence within less than 10 iterations.

\begin{figure} [ht]
   \begin{center}
   \begin{tabular}{c} 
   \includegraphics[height=5.5cm]
   {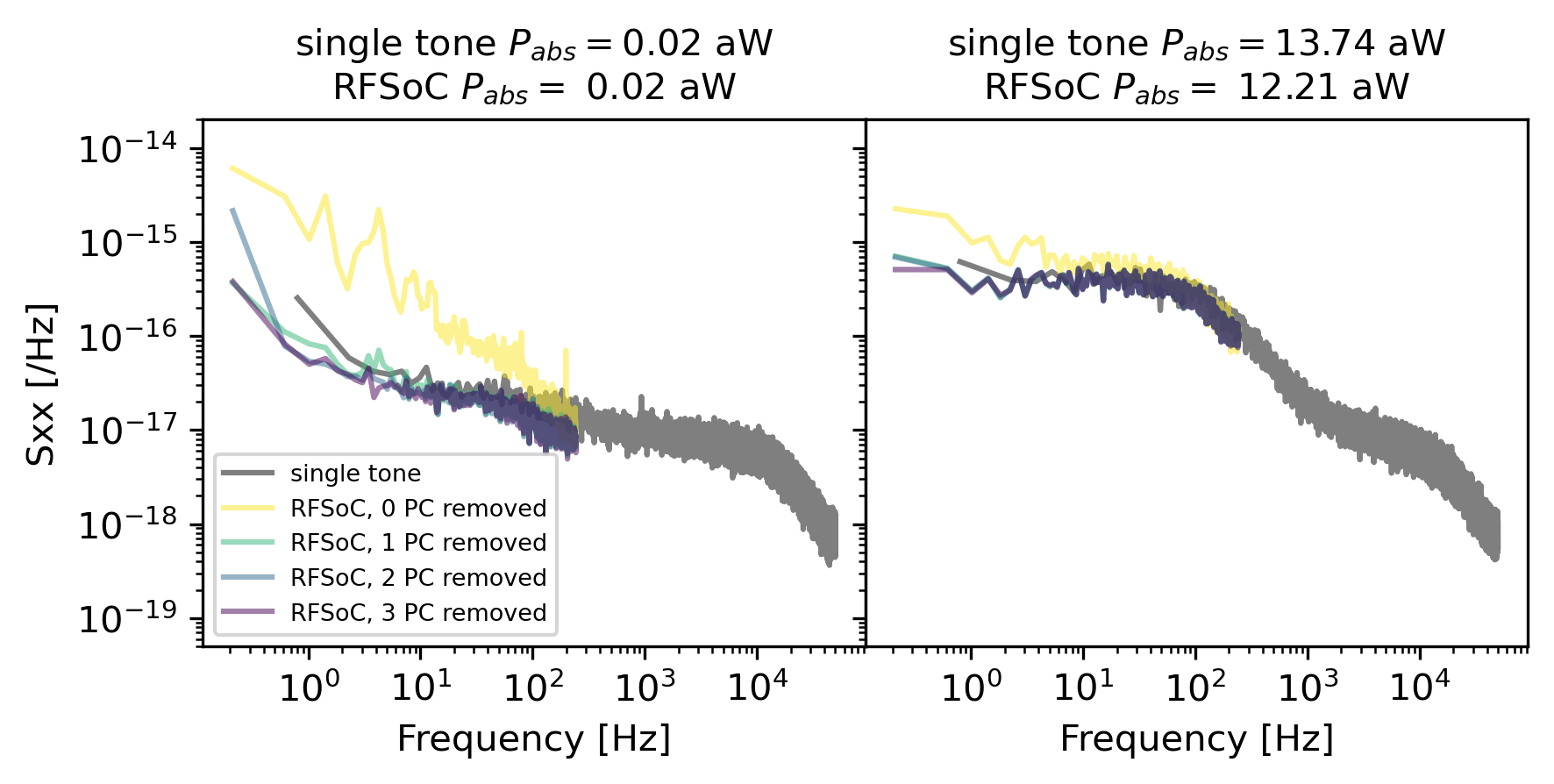}
   \end{tabular}
   \end{center}
   \caption[example] 
   { \label{fig:pca}
\textit{Left:} $S_{xx}$ measurement of a single resonator with the blackbody off. \textit{Right:} $S_{xx}$ measurements of a single resonator at an absorbed optical power of 13 aW. Single-tone and multi-tone measurements are both plotted, as well as multi-tone measurements with 1, 2, and 3 principal components removed. The resonant frequency is $f_r=980$~MHz.}
   \end{figure}

After optimizing, we perform a blackbody sweep with the tone powers fixed at the optimized values. At each blackbody power, a transmission sweep and a noise timestream is taken for each detector. After reaching the maximum power, the blackbody is cooled back down and the same measurements are repeated at each power to combat any drifts in resonant frequency or readout frequency. For each detector and blackbody power, the transmission sweep is fit to Equation~(\ref{eq:s21}). After removing the gain and cable delay prefactors, the fit traces out a circle in the IQ plane. The timestream is then converted from DAQ units to an array of phase shifts relative to the center of this circle. To remove common mode noise, the set of all phase timestreams taken simultaneously are factored into principal components using a singular value decomposition, and the eigenvalues of the largest two modes are set to zero. The cleaned phase timestreams are converted to resonant frequency shifts and the power spectral density $S_{xx}$ is calculated. For most detectors, removing one principal component was sufficient to remove the common mode noise, and removing two or three negligibly affected $S_{xx}$ further, so removal of two principal components was chosen for the final analysis result. To validate this analysis method, measurements of a single resonator with $f_r=980$~MHz were taken and the $S_{xx}$ was compared to the RFSoC $S_{xx}$ before and after principal component removal. In Fig.~\ref{fig:pca}, we see rough agreement between the single-tone $S_{xx}$ and the ``cleaned" multitone $S_{xx}$ at both low and high optical powers. We note that the single-tone measurement and the multitone measurements were performed on separate cooldowns, between which the device was removed from the fridge. Thus, we cannot be sure whether the common mode noise is due to the RFSoC or another part of the readout chain, and this is a subject for future investigation.

\begin{figure} [ht]
   \begin{center}
   \begin{tabular}{c} 
   \includegraphics[height=5cm]{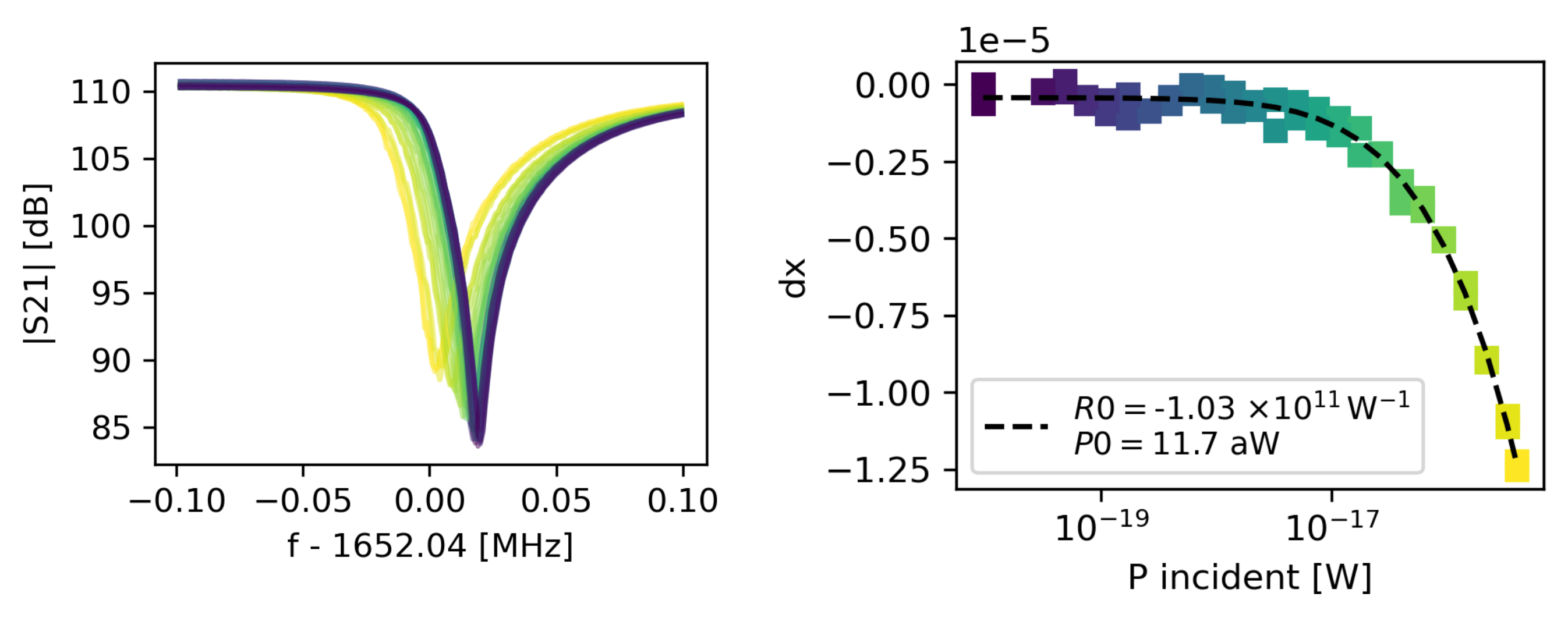}
   \end{tabular}
   \end{center}
   \caption[example] 
   { \label{fig:resResponse} 
\textit{Left:} The transmission curves of a KID throughout the blackbody sweep, with higher power indicated by brighter color. \textit{Right:} The fitted resonant frequencies $f_r$ plotted against incident power on the detector, with a fit to Equation~(\ref{eq:response}). The resonant frequency is $f_r=1652$~MHz.}
   \end{figure}

The fitted resonant frequency values $f_r$ versus blackbody power are used to determine the detector's responsivity. The Mattis-Bardeen theory predicts a constant responsivity at low optical loading $(R_0)$, followed by a decrease at high optical loading $(P>P_0)$ \cite{shd2023}, as shown in the left side of Equation~(\ref{eq:response}). To obtain values for $R_0$ and $P_0$ for each detector, we fit to the integrated response equation in the right side of Equation~(\ref{eq:response}). An example of this fit is shown in Fig.~\ref{fig:resResponse}.

\begin{equation}
    \label{eq:response}
    \frac{dx}{dP} = R_0\left[1+\frac{P}{P_0}\right],\quad x(P)-x(0) = 2R_0P_0\left[\left(1+\frac{P}{P_0}\right)^{1/2}-1\right].
\end{equation}

The NEP (referred to units of incident power) is calculated as $\mathrm{NEP}_{inc}=S_{xx}^{1/2} (dx/dP_{inc})^{-1}$. To calibrate out any system optical efficiencies, we fit the $\mathrm{NEP}_{inc}$ values at a noise frequency of 20 Hz to a noise model given by Equation~(\ref{eq:NEPmodel}). This model includes a power-independent term representing detector noise plus a power-dependent term capturing the effect of photon shot noise and recombination shot noise from optically-generated quasiparticles.

\begin{equation}
    \label{eq:NEPmodel}
    \mathrm{NEP}_{inc} = \mathrm{NEP}_{abs}\,\eta_{opt}^{-1}= \left[2h\nu\eta_{opt}P_{inc}\left(1+n_0+\frac{2\Delta_0}{h\nu\eta_{pb}}\right) + \mathrm{NEP}_0^2\right]^{1/2} \eta_{opt}^{-1}.
\end{equation}

The fit parameters are the optical efficiency $\eta_{opt}$ and the detector NEP referred to absorbed power units, $\mathrm{NEP}_0$. The center frequency of the filter stack is $\nu=1.42$ THz, and the photon occupation number $n_0$ at the detector is calculated to be negligible at our blackbody temperatures and expected optical efficiency. To determine $\Delta_0$, a bath temperature sweep at the lowest incident power was performed and transmission vs. frequency sweeps were taken and fitted to Equation~(\ref{eq:s21}). The change in resonant frequency $f_r$ was then fitted to Equation~(\ref{eq:thermal_fr}). 

\begin{equation}
    \label{eq:thermal_fr}
    f_r(T) = f_{r,0}\left(1+\frac{F\delta_0}{\pi}\left[\mathrm{Re}\left(\frac{1}{2}-\frac{1}{2\pi j}\frac{hf_{r,0}}{k_BT}\right)-\ln\frac{hf_{r,0}}{k_B T}\right]-\alpha\gamma S_2(f_{r,0}, \Delta_0, T) \frac{n_{th}(T)}{4N_0 \Delta_0}\right).
\end{equation}

\begin{figure} [ht]
   \begin{center}
   \begin{tabular}{c} 
   \includegraphics[height=5.2cm]{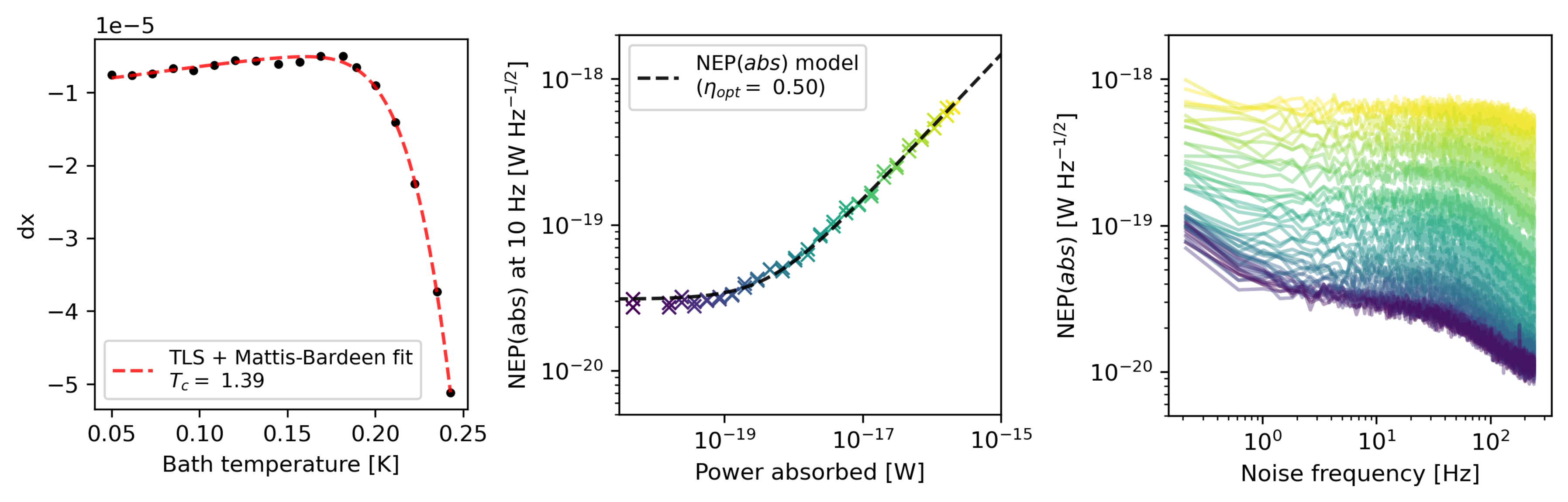}
   \end{tabular}
   \end{center}
   \caption[example] 
   { \label{fig:resNEP} 
Plots illustrating the analysis method on a resonator with resonant frequency $f_r=1652$~MHz. \textit{Left:} Temperature-induced frequency shift of a resonator, with a fit to Equation~(\ref{eq:thermal_fr}). \textit{Center:} Plot of NEP$_{abs}$ vs. power absorbed at a noise frequency of 10Hz, with a fit to Equation~(\ref{eq:NEPmodel}). \textit{Right:} Curves of NEP$_{abs}=\eta_{opt}\,\mathrm{NEP}_{inc}$ versus noise frequency at each blackbody power.}
   \end{figure}

The second term in parentheses of Equation~(\ref{eq:thermal_fr}) captures the TLS frequency shift, and the third term captures the Mattis-Bardeen shift from thermal quasiparticles. The fit parameters are $F\delta_0$, $f_{r,0}$, and $\Delta_0$; we take $\gamma=1$ for the thin-film limit, $\alpha=0.95$ based on electromagnetic simulations of the absorber, $N_0 = 1.72\times 10^{10}\,\mathrm{eV}^{-1}\,\mu\mathrm{m}^{-3}$ for the single-spin density of states at the Fermi level for aluminum, and use standard expressions for $S_2$ and $n_{th}$ \cite{jz12, devisser_thesis}. The superconducting gap fits across the array were sharply peaked around $\Delta_0=0.211$~meV ($T_c=1.39$~K). The pair-breaking efficiency $\eta_{pb}$ is assumed to be 0.5 based on the results of Hailey-Dunsheath et. al\cite{shd2023}. We note that the value of the recombination term with these values is $2\Delta_0/(h\nu\eta_{pb}) = 0.14$, which is much less than 1 and thus the precise values of $\Delta_0$ and $\eta_{opt}$ will only weakly impact the fit. Rather than using the fitted NEP$_0$ value as the result for the detector NEP after fitting to Equation~(\ref{eq:NEPmodel}), we calculate the detector NEP directly from our measured NEP$_{inc}$ values as NEP$_{det}=\eta_{opt}\,\overline{\mathrm{NEP}}_{inc}$, where $\overline{\mathrm{NEP}}_{inc}$ is the mean of our measured NEP$_{inc}$ values at incident powers below 0.1~aW. Examples of fits to Equation~(\ref{eq:NEPmodel}) and Equation~(\ref{eq:thermal_fr}) are shown in Fig.~\ref{fig:resNEP}.

\section{results}
\label{sec:results}

\begin{figure} [ht]
   \begin{center}
   \begin{tabular}{c} 
   \includegraphics[height=5.5cm]{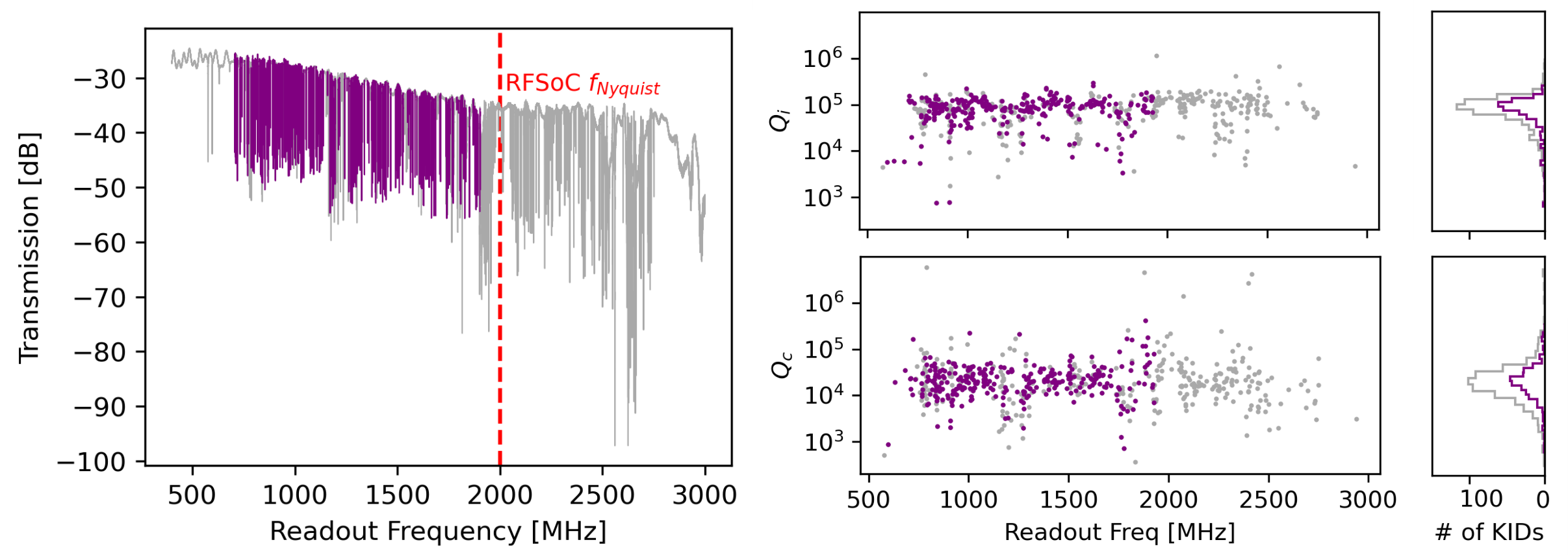}
   \end{tabular}
   \end{center}
   \caption[example] 
   { \label{fig:iqfits} 
\textit{Left:} A full transmission sweep of the device using the single-tone system, with resonances included in the NEP measurements highlighted in purple. \textit{Top Center:} Plot of the fitted internal quality factors $Q_i$ versus frequency. \textit{Top Right:} Histogram of fitted $Q_i$ values. \textit{Bottom Center:} Plot of the fitted coupling quality factors $Q_c$ versus frequency. \textit{Bottom Right:} Histogram of fitted $Q_c$ values.}
   \end{figure}

From the single-tone transmission sweep of the device, 585 resonance features were identified. We note that previous arrays fabricated for PRIMA have demonstrated $>90\%$ yield \cite{footeltd20,chrisSPIE}, and the cause of the decrease in yield for this array is currently under investigation. 569/585 of the resonant frequencies were within the target readout bandwidth of 500~MHz to 2400~MHz, with the 16 remaining resonance features lying above 2400~MHz. The median coupling quality factors and internal quality factors are $Q_c=19,300$ and $Q_i=86,000$. Of the 585 resonances, 436 had frequencies below 1.9 GHz and were read out with the RFSoC. Of these, 295 were selected for the final analysis result. The other KIDs were rejected due to a poor fit to either Equation~(\ref{eq:response}) or Equation~(\ref{eq:NEPmodel}). In Fig.~\ref{fig:iqfits}, the 295 KIDs are highlighted in purple. The values of $Q_c$ and $Q_i$ for the pixels with NEP measurements have the same mean and standard deviation as the full array, suggesting that this subset is a good representation of the full array.

\begin{figure} [ht]
   \begin{center}
   \begin{tabular}{c} 
   \includegraphics[height=4.5cm]{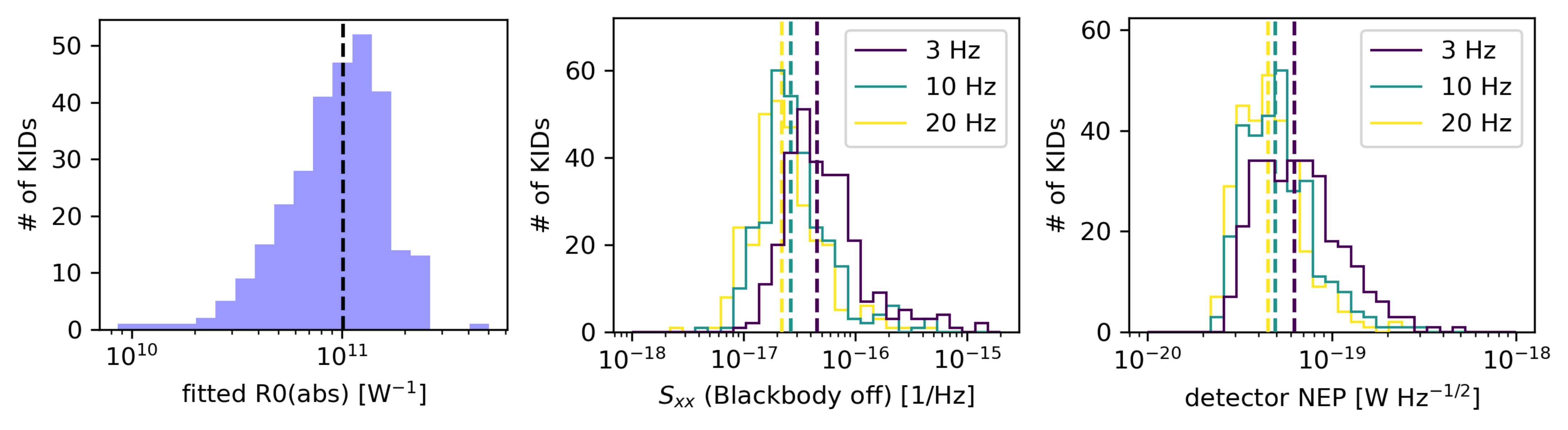}
   \end{tabular}
   \end{center}
   \caption[example]{\label{fig:NEPhist} \textit{Left:} Histogram of responsivities referred to units of absorbed power. \textit{Center:} Histograms of common-mode cleaned $S_{xx}$ values at noise frequencies of 3~Hz, 10~Hz, and 20~Hz. \textit{Right:} Histograms of detector NEP values at noise frequencies of 3~Hz, 10~Hz, and 20~Hz. }
   \end{figure}

Using the optical efficiencies $\eta_{opt}$ extracted from the NEP model fits, we calculated the detector responsivities in units of absorbed power as $R_0(abs) = R_0/\eta_{opt}$ using the fitted values of $R_0$ with respect to incident power units. This yielded a median value of $1.02 \times 10^{11} \mathrm{W}^{-1}$. A histogram of $R_0(abs)$ is plotted in Fig.~\ref{fig:NEPhist} for the analyzed KIDs. Also plotted are dark $S_{xx}$ and detector NEP values. At a noise frequency of 3 Hz, the mean detector NEP is $7.7 \times 10^{-20}$ W Hz$^{-1/2}$, and $79\%$ of the KIDs are below the goal NEP of $1\times 10^{-19}$ W Hz$^{-1/2}$. At a noise frequency of 10 Hz these figures are $5.8 \times 10^{-20}$ W Hz$^{-1/2}$ and $92\%$.

\section{Discussion}

We have presented multitone measurements of a flight-like array designed to observe at 210~$\mu$m for PRIMA's FIRESS instrument, obtaining NEP measurements for $295/436=68\%$ of KIDs with readout frequencies accessible by the RFSoC. Of these KIDs, $92\%$ demonstrated NEPs below the $1\times 10^{-19}$~W~Hz$^{-1/2}$ goal at a noise frequency of 10~Hz, which is the baseline modulation frequency for PRIMA.

Future work will involve investigation of the source of the common mode noise observed in the multitone measurements. We expect much of the $1/f$ drift in the system to come from the internal clock chip, which is a known issue with our RFSoC. If this hypothesis is true, we can overcome the drift with the use of an external clock reference.

We will also work to further increase the readout yield. We found that poor fits to Equation~(\ref{eq:response}) or Equation~(\ref{eq:NEPmodel}) were caused by resonator frequency collisions, misplaced readout tone frequencies, or readout tone power above the resonator bifurcation power. Resonators that were read out at too high a readout power can be recovered by starting the optimization at a lower readout power and increasing the power in smaller increments. Resonators where the readout tone ended up off-resonance can be recovered by more carefully tracking the resonance feature between the initial frequency scan of the device and start of the optimization routine. Colliding resonances can be addressed in the short term by fitting to a more general form of Equation~(\ref{eq:s21}) that includes two or more resonance features. In the long term, colliding resonances can be fixed by lithographically etching away IDC tines once the resonant frequencies are mapped to spatial locations on the wafer. An LED mapping scheme has been used to robustly identify the spatial locations of KIDs for the TIM experiment \cite{simonSPIE}, and work is ongoing to produce such a spatial map for a flight-like PRIMA array \cite{chrisSPIE}.  

\acknowledgments 
 
The research was carried out at the Jet Propulsion Laboratory, California Institute of Technology, under a contract with the National Aeronautics and Space Administration (80NM0018D0004). This work was funded by the NASA (Award No. 141108.04.02.01.36)—to Dr. C. M. Bradford.

\bibliography{report} 
\bibliographystyle{spiebib} 

\end{document}